# Effect of encapsulated graphene oxide on alginate-based bead adsorption to remove acridine orange from aqueous solutions

**Ling Sun[*], Fugetsu Bunshi**

Laboratory of Environmental Remediation, Graduate School of Environmental Science, Hokkaido University, Sapporo 060-0810, Japan

---

E-mail: captainsun@ees.hokudai.ac.jp (L. SUN)

**Abstract**

Environmentally-benign high-performance graphene oxide (GO)/alginate-based absorbents were prepared to eliminate acridine orange selected as a typical dye. Characterizations demonstrated GO well encapsulated and its promotion of pore formation on structure. Kinetic studies exhibited that the addition of GO shortened the adsorption equilibrium time, raised the initial rate and the adsorption capacity. Isotherm studies indicated the adsorptive behavior followed Langmuir type, and higher maximum capacity was obtained in the presence of GO. The adsorption positively responded to pH increased from acidic to weakly alkaline. At low pH, GO would contribute dominantly to the adsorption, whereas alginate component was inhibited.

**Keywords**

GO, Alginate beads, Acridine orange, Ion exchange

# 1. Introduction

Graphene oxide (GO) in nature has large specific surface area and high hydrophilicity. And there could produce a wide series of interactions between the intrinsic functional groups implanted on GO and contaminants. These properties enable GO highly desirable for applications in water remediation. It has demonstrated, few-layered GO had superior adsorptive capability to eliminate various heavy metal ions, such as $Cd^{2+}$, $Co^{2+}$, and $Pb^{2+}$ [1, 2], where the pH-dependent electrostatic attractions among the oppositely charged counter ion/molecule accounted for the results. Besides, GO could reach a theoretical maximum capacity of 313 mg/g for the removal of tetracycline antibiotics from the aqueous solution, via pi–pi interaction and cation–pi bonding [3]. Furthermore, GO is also highly applicable in other ways, like catalyst-supporter or graphene precursor, etc. Nanoflower-like CuO-decorated GO was applied in the photocatalytic degradation of dyes [4]. A magnetic chitosan/GO composite was prepared for the efficient adsorption of methylene blue [5]. One-pot solvothermal synthesized magnetite/reduced GO was used to effectively treat industrial wastewater and lake water [6].

However, an increasing fear is spreading throughout the communities towards GO/graphene's hazards on health during the production, utilization, and disposal, once uptaken by living beings [7, 8]. But common awareness has not been well established. Ahead of that, if necessary, some protective measures are welcome to avoid the possible environmental hazards. Immobilization/encapsulation of GO deserves as an option. Alginate, a heteropolysaccharide composed of β-D-mannuronic acid and α-L-guluronic acid, widely distributes in diverse seaweeds and bacteria. It forms stable gels that could be cross-linked by various di/trivalent cations, such as $Ba^{2+}$, $Ca^{2+}$, and $Fe^{3+}$. Of importance, these gels feature biosafety and easy processing. Entrapment using alginate beads has been extensively investigated for the uptake/separation towards various metal ions retained in water [9-13] as well as some dyes [14, 15], and further indicated high selectivity for water treatment. For example, divalent calcium/trivalent ferric ion cross-linked alginate gel entrapping active carbon powder indicated selective adsorption among differently charged model compounds, such as negatively charged humic acid, neutral *p*-chlorophenol, and positively charged methyl blue [16]. In addition, carbon nanotubes-encapsulated barium cross-linked alginate beads had applied to the selective removal of ionic dyes, such as acridine orange, ethidium bromide, eosin bluish, and orange G, and also had nice biocompatibility [17]. Hydrophilic and flexible alginate/polyurethane composite foams with improved strength could multiply their regenerated use for highly selective binding to $Pb^{2+}$ [18].

Few reports had included GO incorporated with alginate matrix for decontamination use, except for an example of GO/alginate fiber using GO as strengthening element [19]. Previously we have demonstrated GO was highly efficient (~3.3g $g^{-1}$) toward a cationic dye, acridine orange (AO), especially further with an simultaneous in-situ reduction route [20]. Herein from the viewpoint of environmental friendliness, GO was incorporated into alginate-based matrixes to prepare novel adsorbents (calcium alginate beads and porous alginic beads), and we would like to further reveal the possibility over the effectiveness of removal of cationic dye AO from aqueous solutions. Meanwhile, characterizations of materials and relevant studies over the adsorptions process and mechanism were also discussed.

## 2. Experimental

### 2.1. Materials and reagents

Graphite (Commercial Number EC1000, average particle size 15 μm as manufactured) was purchased from Ito Kokuen Co., Ltd, Mieken, Japan. Other chemicals such as sodium alginate, calcium carbonate, and calcium chloride were obtained from Wako Pure Chemical Industries, Ltd., or Sigma-Aldrich Inc., Japan, unless specifically noted.

### 2.2. Preparation of GO-encapsulating alginate beads

Sodium alginate (500 ~ 600 mPa s with a concentration of 10 g L$^{-1}$ at 293 K as marked) was dissolved into deionized water at a mass fraction of 2 % before use. A modified Hummers-Offeman method[21] was applied to prepare GO, of which the weight percent was about 0.10 %.

Calcium alginate beads are commonly prepared by a $CaCl_2$-hardening method [17, 22]. To prepare conventional GO/alginate beads (denoted as SA-GO-N), in brief, 50 g of GO solution was homogeneously mixed with 250 g of sodium alginate solution. A self-made apparatus comprised a 500 ml container for storing the raw solution, an airflow controller and an air pump. With a given internal pressure, the above solution was dripped continuously through the nozzle into a magnetically stirred $CaCl_2$ solution (weight percent about 6 %). Then beads were thoroughly rinsed several times using a 100 μm mesh sieve. Also, calcium alginate beads without GO, denoted with Pure SA-N, were also prepared for comparison.

To prepare porous GO/alginic beads, a modified method was applied with HCl (weight percent 5 %) instead of $CaCl_2$. Briefly, calcium carbonate was ground using a mill system for 48 hours to decrease the particle size. Then, a GO solution of about 50 g and calcium carbon of about 2.5 g were fully mixed. This was followed by the addition of 250 g of sodium alginate and a subsequent mixing. A homogeneous calcium carbonate/GO/sodium alginate solution was readily prepared as the raw solution. Similarly to the above, the alginic beads were by the drop-wise method in HCl. During such, $CO_2$ bubbles were also produced during the reaction between $CaCO_3$ and HCl, and were also caged in the beads to form macro pores. These beads were denoted as SA-GO-M. For comparison, bubble-containing alginate beads without GO were also prepared, denoted as Pure SA-M. All beads were preserved in a fridge. The dry weight of each kind of beads was obtained from averaging over 60 beads by drying beads in a 353 K oven until the weight became constant.

### 2.3. Adsorption

Time-dependent adsorptions were performed at room temperature (293 ± 2 K). Beads of around 6 mg dry weight were mixed with AO of determined concentration (50mL, 20mg L$^{-1}$) and given a 300 rpm shake for some time until the adsorption reached their equilibrium. The residual of AO was determined by a UV-Vis spectrometer (JASCO V-570 spectrophotometer). Isothermal experiments were also performed at room temperature (293 ± 2 K) to describe the relationship between the equilibrium concentration of AO and the corresponding equilibrium capacity of beads. Each group in all experiment above set in triplicate for determination.

The initial concentration as well as that at each pre-set interval, and at the equilibrium were respectively denoted as $C_0$ (g L$^{-1}$), $C_t$ (g L$^{-1}$), and $C_e$ (g L$^{-1}$), respectively. The adsorption capacity ($q_t$, g g$^{-1}$) was calculated by the equation

$$q_t = \frac{V \times (C_0 - C_t)}{m} \quad (1)$$

where $V$ (mL) represents the volume of the suspension, and $m$ (g) is as described as the dry weight of the sorbent we used in the system.

The equilibrium capacity ($q_e$) was obtained when the adsorption was at the equilibrium $C_e$.

### 2.4. Material characterization

GO samples were characterized by multiple ways. Diameter of the beads was recorded by using shape-recognition software over 60 beads. Atomic force microscopy (AFM) used an Agilent series 5500 AFM instrument in tapping mode at a scanning rate of 0.5 Hz. Fourier transform infrared spectroscopy (FTIR, FT/IR-6100 FT-IR Spectrometer, JASCO) was used in an attenuated total reflection mode (ATR). Scanning electro-microscopy (SEM, JSM-6300, JOEL), Thermogravimetric analysis (TGA, TG/DTA 6200, SII Exstar6000, with a heating rate of 5 ℃ per minute under a N$_2$ atmosphere) and X-ray diffraction (XRD) using RINT 2000 (Rigaku Denki, Ltd, X-ray $\lambda_{Cu\ k\alpha}$= 0.154 nm) were also applied. The specific surface area (SSA) was obtained by aBrunauer-Emmet-Teller method from N$_2$ sorption at 77 K (Yuasa Ionics Autosorb-6). All samples were degassed at 80 ℃ for 2 h prior to determination.

### 2.5. Data analysis

#### 2.5.1 Adsorption kinetics

To study the sorption kinetics, the pseudo-first-order and pseudo-second-order rate models were introduced. The pseudo-first-order rate equation of Lagergren is as follows [23, 24]:

$$\frac{dq_t}{dt} = K_1(q_e - q_t) \quad (2)$$

It can be expressed in linear form:

$$\ln(q_e - q_t) = \ln q_e - K_1 t \quad (3)$$

where $q_e$ and $q_t$ represent the adsorption capacities at equilibrium and as time sampled at every interval for the beads (g g$^{-1}$), respectively; $K_1$ is the rate constant of the pseudo-first-order adsorption (min$^{-1}$).

For the pseudo-second-order rate model [24], the equation is given in differential form as:

$$\frac{dq_t}{dt} = K_2(q_e - q_t)^2 \quad (4)$$

It is then converted as follows:

$$\frac{t}{q_t} = \frac{1}{K_2 q_e^2} + \frac{t}{q_e} \quad (5)$$

where $q_e$, $q_t$ represent the adsorption capacity at equilibrium and at every interval (mg g$^{-1}$), respectively; $K_2$ is the rate constant of adsorption for the pseudo-second-order model (g g$^{-1}$ min$^{-1}$); $K_2 q_e^2$ is the initial adsorption rate (mg g$^{-1}$ min$^{-1}$).

Web-Morris model was used to describe the pore diffusion [25, 26]. It depends on the most solute uptake varying almost proportionally with the evolution of contact time, which is written as follows:

$$q_t = K_{int} t^{1/2} \quad (6)$$

where $K_{int}$ is the pore diffusion rate constant (mg g$^{-1}$ min$^{0.5}$).

*2.5.2 Adsorption isotherms*

Langmuir and Freundlich isotherms were commonly used by their linearized forms presented as follows [20]:

$$\frac{1}{q_e} = \frac{1}{q_m} + \frac{1}{K_L q_m c_e} \quad (7)$$

$$\ln q_e = \ln K_F + \frac{1}{n} \ln c_e \quad (8)$$

Where $q_e$ is the adsorption capacity at equilibrium (g g$^{-1}$); $c_e$ is the equilibrium concentration (g L$^{-1}$); $q_m$ is the theoretical monolayer maximum capacity of adsorbent (g g$^{-1}$); $K_L$ is the Langmuir adsorption constant related to the energy of adsorption (L g$^{-1}$); $K_F$ is the Freundlich adsorption constant related to the energy of adsorption (L g$^{-1}$), and $\frac{1}{n}$ is the dimensionless heterogeneity factor.

## 3. Results and discussion

*3.1. Characterization*

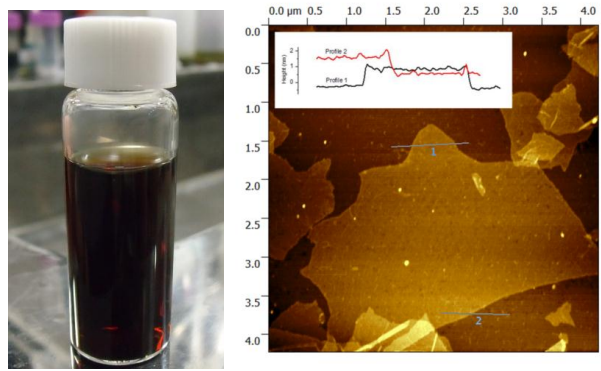

Fig.1. (Left) Photographic image of as-prepared GO solution; (Right) a typical AFM image of GO deposited on freshly exfoliated mica, where the inset profiles indicate the GO sheet thickness.

Bulk solutions (left in Fig. 1) containing fully exfoliated GO were produced from commercially available expanded graphite. From a typical AFM image (right in Fig. 1), GO sheets were micrometer long, and about 1 nm thick *vs.* 0.35 nm for graphene owing to the implantation of oxygen-containing functionalities onto both surfaces.

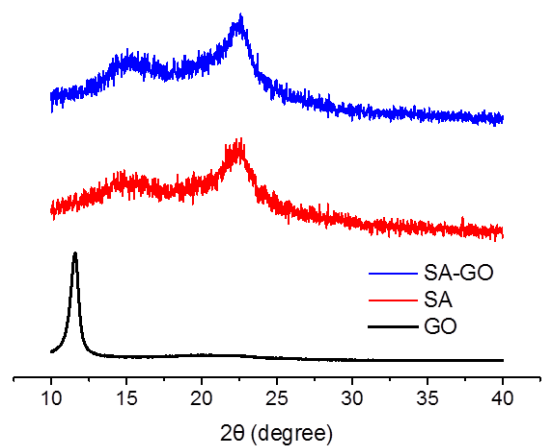

Fig. 2. XRD pattern of alginate/GO film (SA-GO), alginate film (SA) and GO film.

The XRD pattern of GO, sodium alginate and their hybrid film were shown in Fig. 2. GO had a typical peak around 11.6 ° corresponding to interlayer space around 7.6 Å. It had no distinct difference for alginate before and after GO addition, compared with GO. There exist two typical peaks around 14 and 22.5° of 2θ which are related to the lateral packing among molecular chains and the layer spacing along the molecular chain direction [27], respectively. Of note, a small reduction in d-spacing value from 6.01 Å (SA) to 5.74 Å (SA-GO) indicated an interaction between the carboxyl groups of sodium alginate with hydroxyl groups of GO [28]. That is, the water-soluble GO would maintained well in the sodium alginate solution, so that solutions of GO and alginate were prepared homogenously.

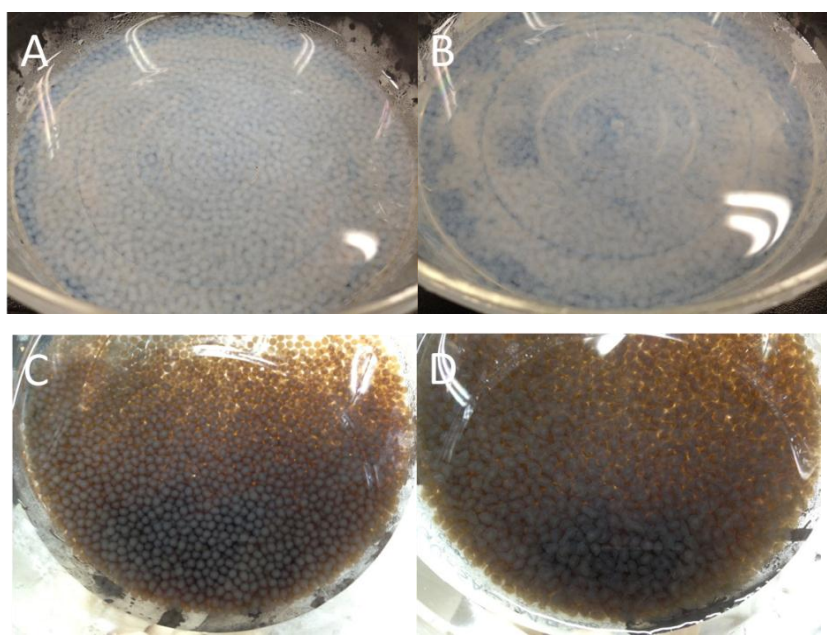

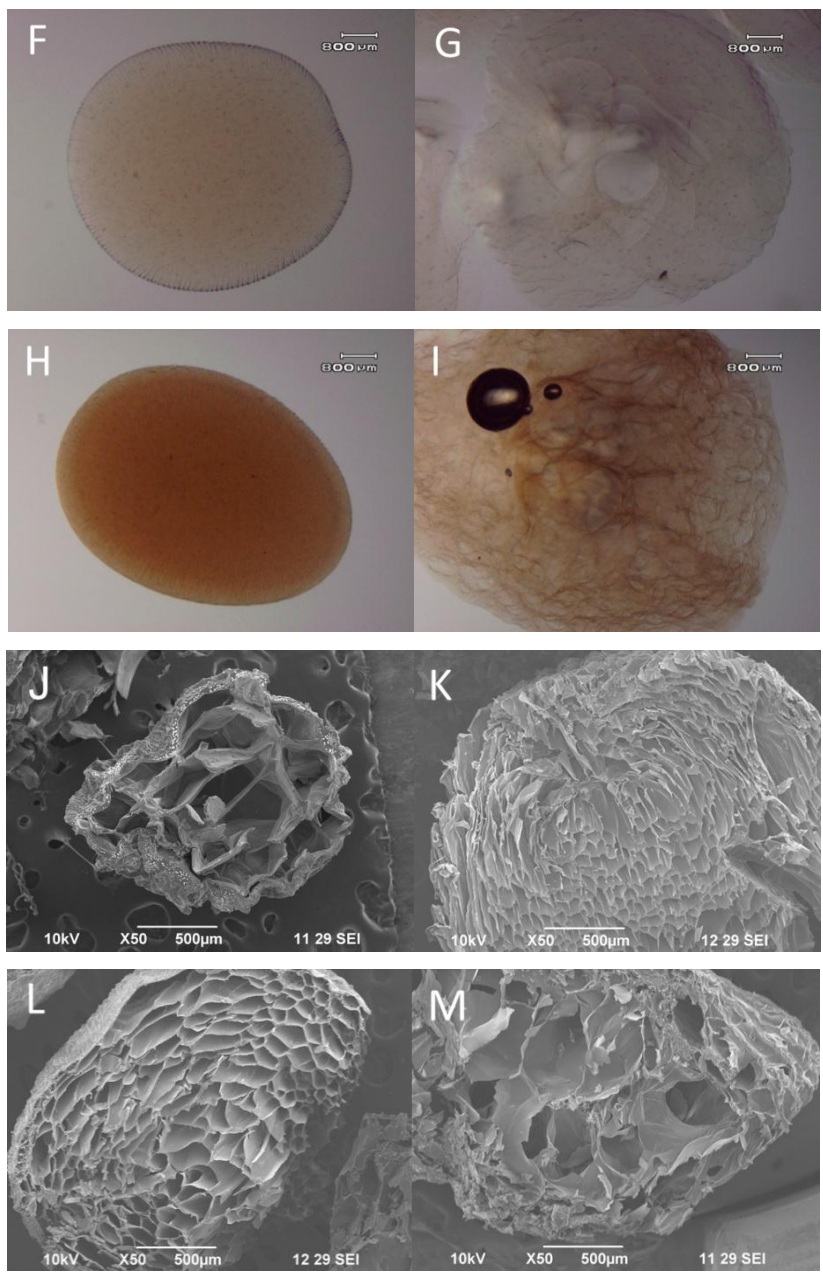
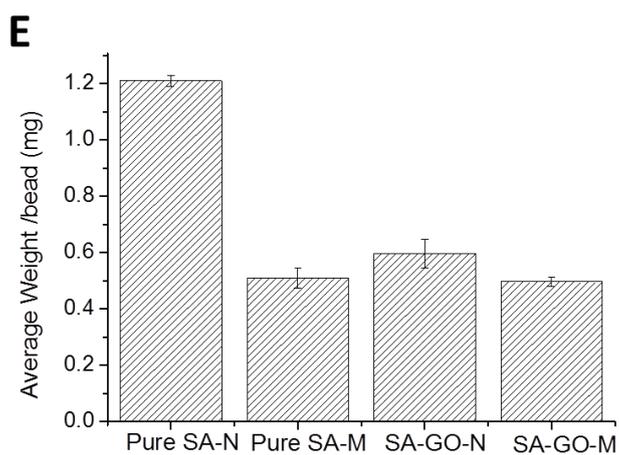

Fig. 3. Photographic and SEM images of beads: (a, f, j) Pure SA-N, (b, g, k) Pure SA-M, (c, h, l) SA-GO-N, and (d, i, m) SA-GO-M. Histogram (e) of their dry weights obtained by averaging 60 beads dewatered in a 353 K oven.

To prepare porous alginic beads, as-added calcium carbonate would prefer more homogeneous distribution before reaction with hydrochloric acid. It had been mechanically milled so that particles of finer sizes were obtained, more favorable for fast dispersion in the alginate solution [29]. The milling reduced the particle size (size > 3 μm, from nearly 100 % to less than 25 %). In addition, sodium alginate is reportedly capable to mono-disperse water-insoluble materials, i.e., MCNTs [30]. However, the case for calcium carbonate dispersion by sodium alginate of 2.5 % (w/w) remained unsatisfying, even with a mechanical stirring and extra ultra-sonication. It's time-consuming and undesirable for scalable production. In the meantime GO has also been used for dispersing insoluble materials, i.e., graphite and CNT as a molecular dispersing agent [31]. In this case, a relatively stable GO/calcium carbonate suspension has been prepared with a mild shake. Batches of beads were obtained within minutes at a kilogram scale (Fig. 3a-d).

Figs.3f ~ 3m show the microscopic images of freshly-prepared beads and their cross sections of after freeze-drying. Inner structures were different from each other: tight and sphere-like for the calcium beads, and yet fluffy and macro porous for the alginic beads. The size of the fresh beads were in the order: SA-GO-M (5.40±0.55 mm) > Pure SA-M (4.78±0.41 mm) > SA-GO-N (3.64±0.25 mm) > SA-GO-N (3.35±0.24 mm). The SSA values of each adsorbent had the order: SA-GO-M (31.3 $m^2/g$) > SA-GO-N (22.9 $m^2/g$) > Pure SA-M (11.4 $m^2/g$) > Pure SA-N (0.4 $m^2/g$). After freeze drying, Pure SA-N was largely shrunken because of water loss; and the cross section indicated its "cell" structure with walls of about 20 μm thick. In contrast, thinner crisscrossed "walls/channels" existed for the acid gelated beads, resulting from the release and excessive build-up of $CO_2$, etc. A similar structural difference also appeared to the calcium alginate beads with encapsulating GO (SA-GO-N) as compared to the control (Pure SA-N), owing to the GO promotion in pore formation [32]. Additionally, calcium alginate beads (Pure SA-N and SA-GO-N) settled down (Figs. 3a and 3c), while porous alginic beads (Pure SA-M and SA-GO-M) got floating at the solution surface (Figs. 3b and 3d), that is, the latter had a relatively lighter density. Subsequently, the dry weight was measured (Fig. 3e). On one hand, the calcium beads had weight higher than others; on the other, the beads were lighter per one bead than those without encapsulating GO, like SA-GO-N *vs*. Pure SA-N. It is noteworthy for calcium alginate bead (SA-GO-N), a relatively lighter weight accompanying with porous structure and large diameter, provided larger reachable surface for solute than that of the Pure SA-N.

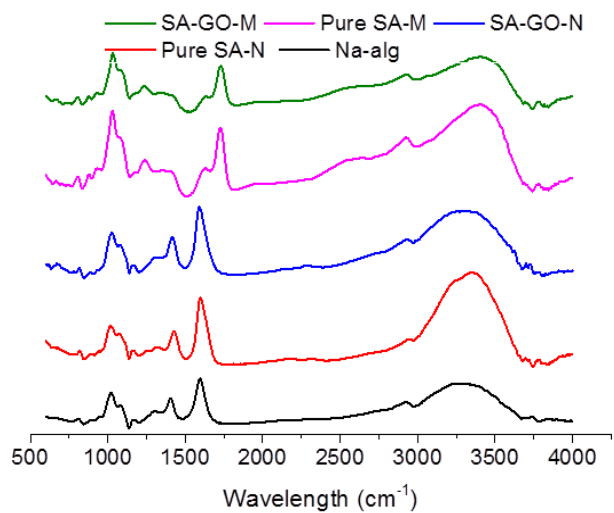

Fig. 4. FTIR spectra of alginate-based beads (Pure SA-N, Pure SA-M, SA-GO-M and SA-GO-N) and sodium alginate.

To clarify the crosslinking process, FTIR was used for analysis. In Fig. 4, broadened peaks at 3398 cm$^{-1}$ were observed indicating of GO interaction with alginate [19]. Historically calcium complexes, such as calcium citrate and calcium carbonate [33, 34], were used as calcium vectors to crosslink alginate through a mild acidification, for example using glacial acetic acid. Here concentrated HCl was instead initially expected to fast release carbon dioxide and calcium ions for internal pore formation and simultaneous cross-linkage of alginate, respectively. Actually the bead composition was affected by the acid. As shown, their spectra were different from calcium alginate. The peaks at 1415 cm$^{-1}$ and 1595 cm$^{-1}$ indicated the symmetric and asymmetric stretching vibrations of carboxylate, respectively, in some metal-alginate interactions [35]. By contrast, a shifted peak at 1728 cm$^{-1}$ represented the typical vibration of carbonyl stretching in alginate acid [35]. For the alginic beads (Pure SA-M and SA-GO-M), as a matter of fact, the pKa of alginate was around 4.2 [36]. The hardening solutions in our case at a lower pH (around 1.9) would cause hydrogen ion gelation, leading to the presence of free carboxylic group, although a large amount of calcium ions fast released and crosslinked the alginate initially. Moreover the strength was not high like calcium alginate, but its adsorption got potentially improved. Previous report had showed an enhanced performance by conversion of calcium alginate fabrics into alginic acid fabrics with additional hydrochloric acid treatment [37]. Besides, alginic "particles" had also been used for collecting rare earth metal from aqueous environments, subject to the intrinsic ion exchange reaction [38].

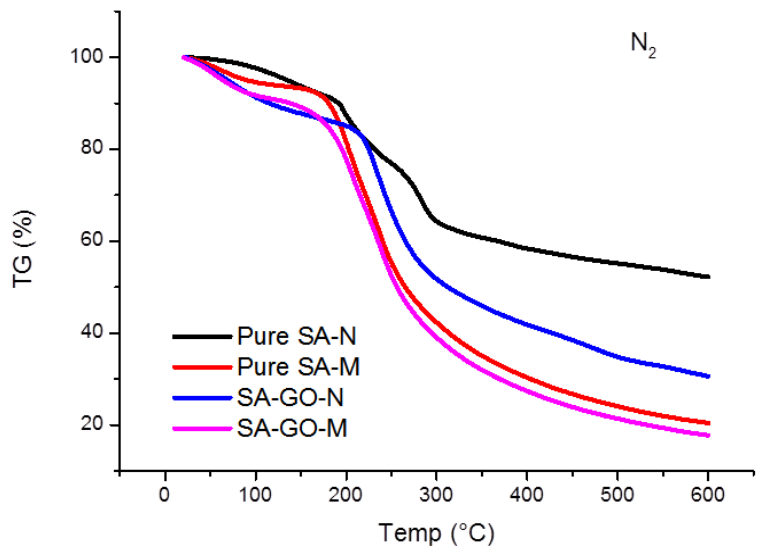

Fig. 5. Thermogravimetric (TG) curves of all beads with a heating rate at 5 K per min under a $N_2$ atmosphere at a flow rate at 90 mL per min.

Further, TG analysis was used to study the thermal stability (Fig. 5). The weight loss below 474 K for calcium alginate beads, 431 K for alginic beads was mostly contributed by the adsorbed and hydrated water. Subsequent was rupture of chains, fragments and monolayers of alginate and decomposition of functional groups of GO [20, 39]. GO was intrinsically instable due to its surface-decorating oxygen functional groups, what rendered beads with less stability. Notably, the residuals were distinctly dependent on the way they were prepared. No clear discrimination of residual was observed for the alginic beads (~20.4 % for Pure SA-N, ~17.7 % for SA-GO-N), but for calcium beads, this discrimination was obvious (~ 52.2 % for Pure SA-N, ~30.6 % for SA-GO-N), indicating a big difference of their structures when GO involved, and this result agreed with the above observed.

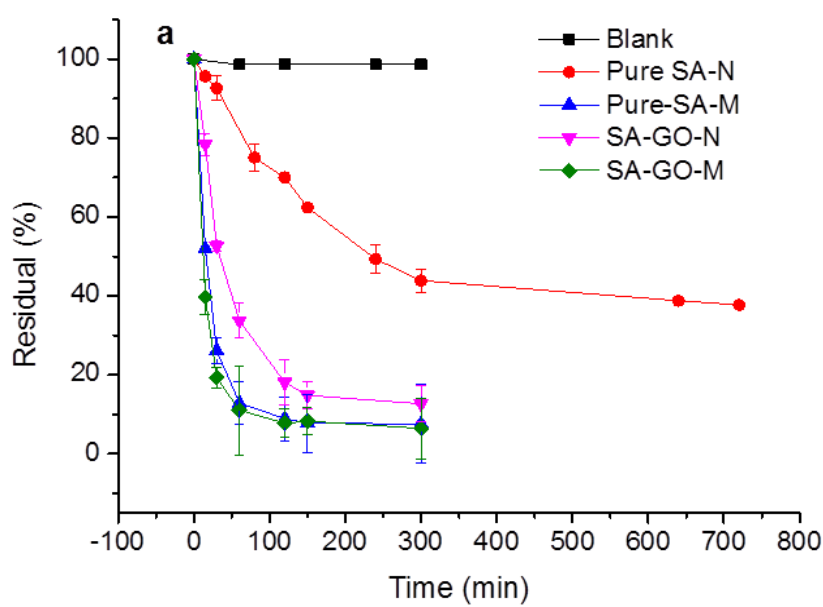

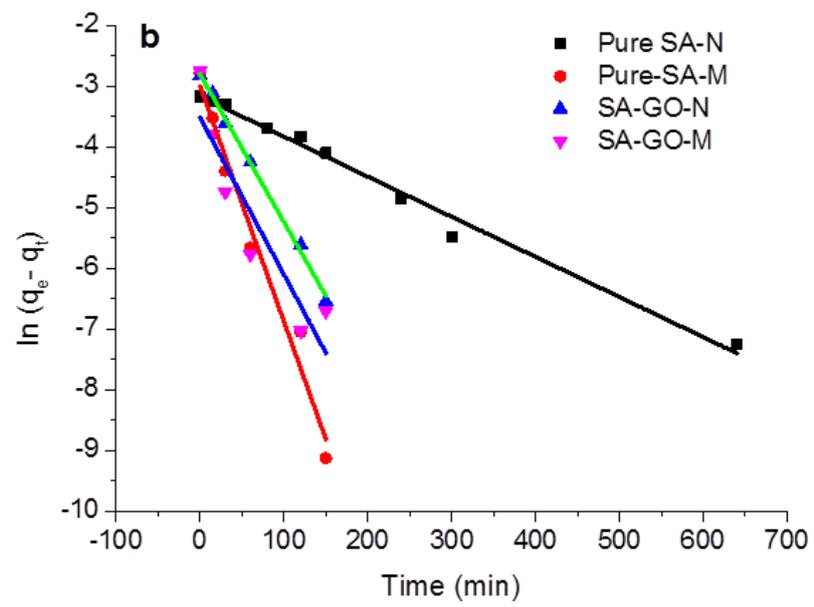

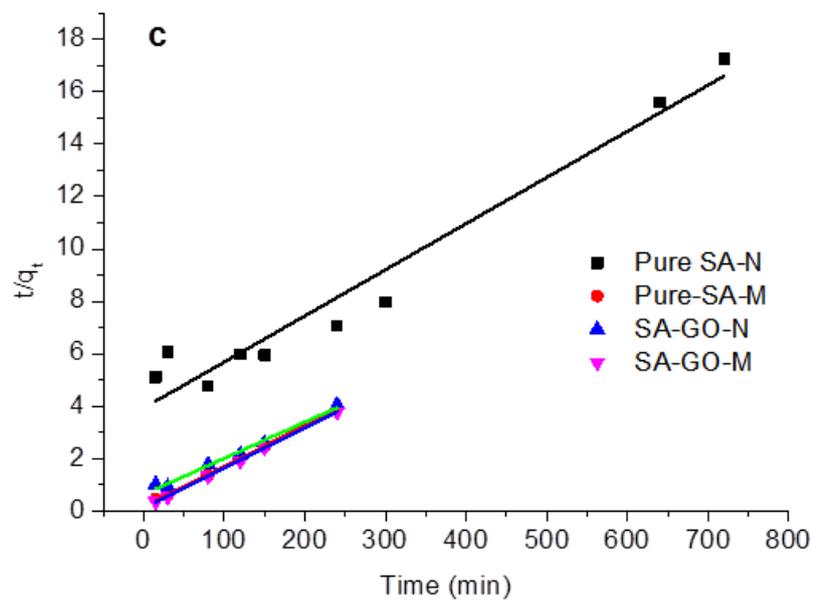

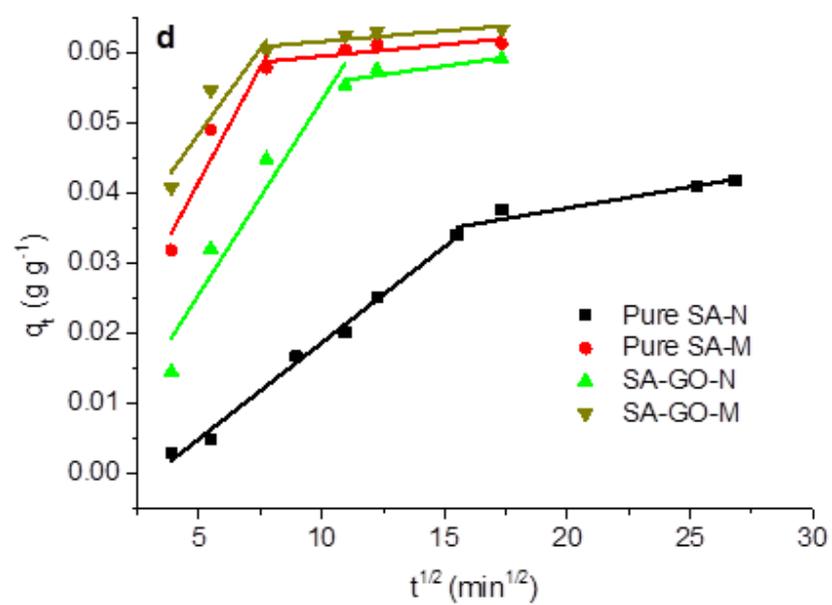

Fig. 6. Time-dependent adsorptions of beads toward 0.02 g L$^{-1}$AO, pH~ 5 (a) and fitted by the pseudo-first-order rate model (b), the pseudo-second-order rate model (c), and Webber's diffusion model (d).

*3.2. Kinetic studies*

Fig. 6 shows the time dependent residual of AO under identical additive amount of beads. The average adsorptive rate was in the order: SA-GO-M (0.21 mg g$^{-1}$ min$^{-1}$) > Pure SA-M (0.20 mg g$^{-1}$ min$^{-1}$) > SA-GO-N (0.19 mg g$^{-1}$ min$^{-1}$) > Pure SA-N (0.07 mg g$^{-1}$ min$^{-1}$). Most adsorptions gradually reached equilibrium with residual less than 15 % at less than 180 min except for the Pure SA-N; The beads became all dark red in color, that is, AO was strongly attracted. Due to the dissociation of carboxyl groups alginate-based beads were negatively charged and showed strong electrostatic interactions to positively charged molecules [16]. Furthermore, for calcium beads, entrapment of GO notably uplifted the removal efficiency from 62.3 % to 87.3 %. But for alginic beads, they were quite similar (92.4% for Pure SA-M; 93.6 % for SA-GO-M). The insignificance showed GO less affected the adsorption of as-prepared porous alginic bead than that of calcium beads. The latter had thick sealed gel structure that suppressed the penetration of molecules deeply into the internal entrapped sorbent [16]. By contrast, dye molecules were able to reach inside through the open-ended channel-like structure of alginic beads.

Tab. 1. Parameters for the pseudo-first-order rate model, the pseudo-second-order rate model and the Weber's pore-diffusion model.

|  | Parameters | Pure SA-N | Pure-SA-M | SA-GO-N | SA-GO-M |
|---|---|---|---|---|---|
| **Experimental** | $q_e$ (mg g$^{-1}$) | 41.8 | 61.3 | 59.0 | 63.3 |
| **Pseudo 1-order** | $R^2$ | 0.9879 | 0.9758 | 0.9972 | 0.8627 |
|  | $K_1$ (min$^{-1}$) | 0.0066 | 0.0387 | 0.0243 | 0.0258 |
|  | $q_{e,cal}$ (mg g$^{-1}$) | 41.9 | 49.6 | 60.4 | 35.2 |
|  | $h_{0,1}$ (mg min$^{-1}$ g$^{-1}$) | 0.277 | 1.920 | 1.468 | 0.908 |
| **Pseudo 2-order** | $R^2$ | 0.9570 | 0.9988 | 0.9868 | 0.9998 |
|  | $K_2$ | 0.0001 | 0.0014 | 0.0003 | 0.0022 |
|  | $q_{e,cal}$ (mg g$^{-1}$) | 56.8 | 64.9 | 72.5 | 65.4 |
|  | $h_{0,2}$ (mg min$^{-1}$ g$^{-1}$) | 0.2544 | 5.8140 | 1.5711 | 9.1996 |
| **Webber's** | $R^2$ | 0.900 | 0.871 | 0.930 | 0.249 |
| **(First segment)** | $K_{inf}$ (mg g$^{-1}$ min$^{0.5}$) | 0.002 | 0.008 | 0.005 | 0.009 |

Pseudo first-order (Eq. 3), pseudo-second order rate (Eq. 5), and Web-Morris pore diffusion (Eq. 6) models were used to study the sorption kinetics (Figs. 6b-d). The suitability was determined by introducing the correlation coefficient ($R^2$). The closer the value $R^2$ to 1, the more applicable the model was. The values of estimated parameters including the characteristic constants were also listed in Tab. 1. Fig. 6b shows all the driving forces $(q_e - q_t)$ decrease over the contact time. In other words, the adsorption rate ($dq/dt$) decreased with time until the equilibrium reached. From Tab. 1, on the one hand, most adsorptions were fitted by both pseudo first-order and pseudo second-order rate models well ($R^2 > 0.9$, except for SA-GO-M with first-order model); on the other, the pseudo first-order model had better fitting over Pure SA-N and SA-GO-N, while the case was reversed for Pure SA-M and SA-GO-M. And correspondingly, the calculated $q_{e,cal}$ were consistent to the experimental data. In addition, higher values for $K_1$ and $K_2$ indicated that the adsorption would take less time to be equilibrium [25]. The GO-encapsulated beads both performed superiorly. Moreover, initial rates of adsorption were also calculated:

$$h_{0,1} = K_1 q_e \quad (9)$$

$$h_{0,2} = K_2 q_e^2 \quad (10)$$

where $h_{0,1}$ and $h_{0,2}$ represent the initial rates (mg min$^{-1}$ g$^{-1}$). As a result the initial rates were in the order: SA-GO-M (9.1) > Pure SA-M (5.8); SA-GO-N (1.5) > Purer SA-N (0.3). Such increases were explainable with the structural changes as well as the increased reactive sites from the GO.

Fig. 6d presents the plot of $q_t$ versus $t^{1/2}$. If the pore diffusion plays as the rate-limiting step, the plot would give a straight line through the origin, with a slope value $K_{int}$, and an intercept that equals to zero. Here all plots were actually multi-linear, consisting of two linear segments. From Tab. 1, the estimated intercepts from regressions were impossibly zero for the first segments for Pure SA-M and SA-GO-M ($R^2 < 0.9$). Thus chemical reaction (or film diffusion), other than the pore diffusion was the rate-limiting factor at their early stage [25, 26]. While for the calcium beads, pore diffusion was the rate-limiting step ($R^2 > 0.9$), and the value related to GO was some larger, showing a faster pore diffusion.

*3.3. Adsorption isotherms*

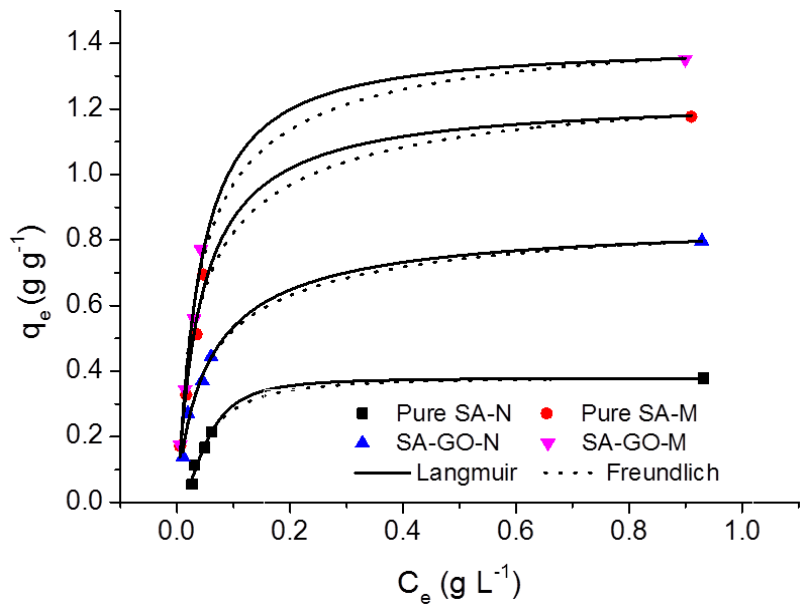

Fig. 7. Adsorption isotherms of AO on beads subsequent fitted by Langmuir (solid lines) and Freundlich (dotted lines) models.

Fig. 7 shows room temperature isotherms of AO absorbed onto beads. A common tendency occurred that the adsorbed amount increased with increase of AO concentration and then gradually reached the plateau maximum. The experimental capacity was in the order: SA-GO-M (1.351 g g$^{-1}$) > Pure SA-M (1.176 g g$^{-1}$), SA-GO-N (0.797 g g$^{-1}$) > Pure SA-N (0.378 g g$^{-1}$) (Tab. 2). With incorporating GO, alginic beads had capacity with a nearly 15 % increase while the capacity for calcium beads had almost doubled.

Tab. 2 Fitting parameters for Langmuir and Freundlich isotherms.

|  | Parameters | Pure SA-N | Pure SA-M | SA-GO-N | SA-GO-M |
|---|---|---|---|---|---|
| Experimental | $q_m$ (g g$^{-1}$) | 0.378 | 1.176 | 0.797 | 1.351 |
| Langmuir | $R^2$ | 0.977 | 0.984 | 0.976 | 0.994 |
|  | $K_L$ (L g$^{-1}$) | 10.867 | 39.147 | 18.261 | 27.092 |
|  | $R_L$ | 9.2E-5* | 2.6E-5* | 5.5E-5* | 3.7E-5* |
|  | $q_{m,cal}$ (g g$^{-1}$) | 0.475 | 0.960 | 0.886 | 1.300 |
| Freundlich | $R^2$ | 0.699 | 0.880 | 0.867 | 0.860 |
|  | 1/n | 0.416 | 0.360 | 0.349 | 0.378 |
|  | $K_F$ (L g$^{-1}$) | 0.451 | 1.479 | 0.950 | 1.735 |

*The concentration was 1000mg L$^{-1}$

To identify the behavior of the absorbate between the liquid-solid phases, typical Langmuir and Freundlich isotherms are used The former is established with assuming the monolayer coverage of solute over specific homogeneous sites within the sorbent [5, 20, 25, 26]; for the latter, it describes the sorption energy exponentially decreases on the complexation of the sorptional centers of an sorbent and the indicated heterogeneous system [39]. As the results in Tab. 2, the adsorptions were better evaluated by the Langmuir model than the Freundlich, for higher coefficients ($R^2 > 0.9$). At the same time, their theoretical maximum capacity also agreed with the corresponding experimental values, showing high uniformity. Therefore the adsorption toward AO in this study followed a monolayer/homogenous behavior. It is noted, the experimental capacity for alginate beads was little lower than the theoretical value; while it was higher, just reversed for the porous alginic beads. Probably an excess uptake of AO occurred because of a dimer or bilayer conformation of AO molecules when the amount of AO adsorbed was over the cationic exchange capacity [40, 41]. Moreover, we could find the high-performance adsorbent with entrapping GO even reached the capacity level as high as those previously reported, for example, the value for SA-GO-M equivalent to the capacity of GO (Tab. 3).

Tab. 3. Literatures regarding adsorptions related to alginate and/or cationic dyes.

| Targeted contaminants | Capacity (mmol g$^{-1}$) | Sorbents | Literatures |
|---|---|---|---|
| **Methylene blue** | 0.052 | Wet CaMB | Rocher V et al. [42] |
| **Methylene blue** | 0.055 | Dry EpiMB-1 | Rocher V et al. [42] |
| **Methylene blue** | 0.14 | GNS/Fe$_3$O$_4$ | Ai L et al.[14] |
| **Methylene blue** | 6.062 | GO | Zhou CJ et al. [43] |
| **p- Chlorophenol** | 1.5 | AG-AC | Lin YB et al. [16] |
| **AO** | 1.2 | Swelling clays | Lv GC et al.[40] |
| **AO** | 4.7 | GO | Previous work [20] |
| **AO** | 1.2 | Pure SA-N | This work |
| **AO** | 2.6 | SA-GO-N | This work |
| **AO** | 3.7 | Pure SA-M | This work |
| **AO** | 4.5 | SA-GO-M | This work |

Another essential characteristic for the Langmuir isotherm is expressed in terms of a dimensionless equilibrium parameter ($R_L$), that is defined as:

$$R_L = \frac{1}{1 + bC_0} \quad (11)$$

where $b$ is the Langmuir constant and $C_0$ is the highest dye concentration (mg L$^{-1}$). The value of $R_L$ indicates the type of the isotherm to be either unfavorable ($R_L > 1$), linear ($R_L = 1$), favorable ($0 < R_L < 1$) or irreversible ($R_L = 0$) [26]. Here the values of $R_L$ were of $10^{-5}$, close to zero, suggesting this process was much favorable, and probably irreversible at extremely high concentration.

## 3.4. pH responsive mechanism

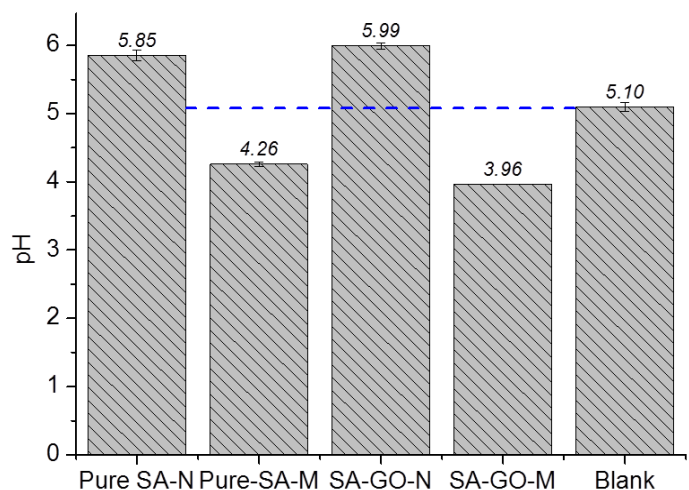

Fig. 8. pH changes of solutions after adsorption in contrary to bulk solutions. AO: 20 mg L$^{-1}$; beads: around 6 mg (dry weight).

Carboxyl groups of alginate made itself intrinsically capable of ion-exchange with positively charged inorganic and organic substances, such as H$^+$, Ca$^{2+}$, Fe$^{3+}$ [16], and methylene blue [15] and this had addressed the mechanism for a majority of adsorptions; thereby it was also preferable for the cationic dye AO adsorbed in our case. Besides, the pH of solution, before and after adsorption, was carefully measured and compared (Fig. 8). The bulk solution of AO (20 mg L$^{-1}$) was a little acidic, with pH around 5. After adsorption, the pH became more lower for the alginic bead adsorption (Pure SA-M and SA-GO-M), due to the release of more hydrogen ions by the way in Eq. 12. By contrast, for alginate beads, released calcium ions eventually decreased the acidity (Eq. 13), and pH increased.

$$\text{Bead} - AO + H^+ \leftrightarrow \text{Bead} - H + AO^+ \quad (12)$$

$$\text{Bead} - AO + Ca^{2+} \leftrightarrow \text{Bead} - Ca + AO^+ \quad (13)$$

Notably, the degree of change on pH differed obviously for the two kinds of beads. The pH preferred more change for the GO-containing beads. In other words, the GO-encapsulated beads were prepared intrinsically with higher adsorptive capacity than that pure alginate or alginic beads. Even more, we further simply immersed the dyed beads into the calcium chloride solution at 5 % (Fig. 9). The solution was instantly colored and this clearly proved a fast reaction upon their contact. The present color indicated the release of AO (judged with UV spectrum), convincingly the result of replacement with calcium ions. However, the control (in water) got no change in the presence of the same beads, even suffered a long-term shake (200 rpm, 72 h).

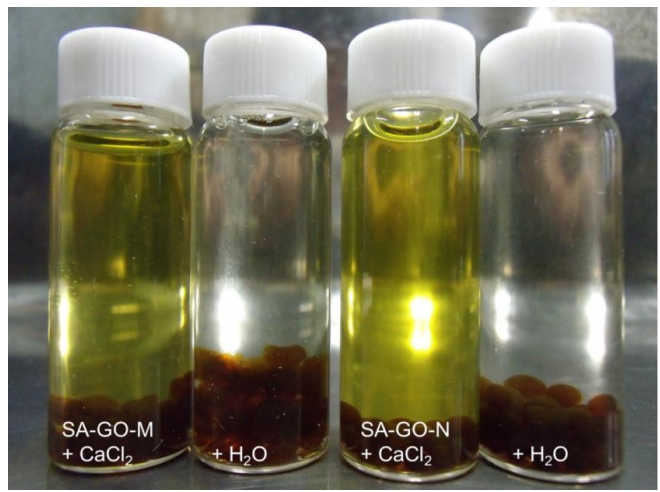

Fig. 9. Photographic image at the moment that dyed beads (SA-GO-M and SA-GO-N) re-immersed in CaCl$_2$ (0.5 M) and deionized water.

*3.5. Effect of solution pH on dye sorption*

In Fig. 10, the effect of pH on adsorption was studied from acid to weakly alkaline, where AO is still protonated since the pKa of AO is around 10.4 [40]. The adsorbent capacity overall increased as pH increased. At low pH, the calcium ions in the beads decreased while carboxylate groups of alginate and GO became progressively protonated [42]. At lower pH, there existed competing adsorption between H$^+$ and cationic dye ions, and the H$^+$ took the priority; at higher pH, more oxygen-containing functional groups were dissociated and therefore through electrostatic attraction, the adsorbent turned more negatively charged, favorable for adsorption [25, 42].

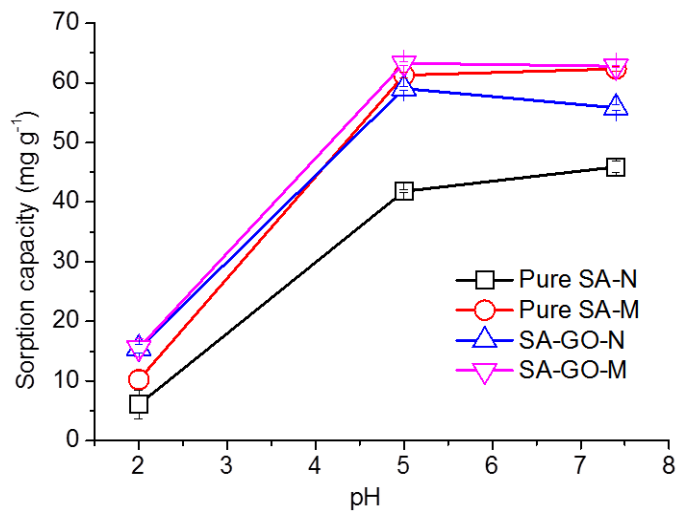

Fig. 10. Adsorptions of dye onto beads under indicated pH range with initial concentration at 0.02 g L$^{-1}$ at room temperature.

Of note, GO functions relying on not only its functional groups, but also the pi-pi electrostatic attraction as well as its amphiphilic affinity. For water decontamination, it mostly features Langmuir-type adsorptive behavior, whenever in the state of being reduced/aggregated [20]. And such behavior also appeared in other cases, like adsorbing $Cu^{2+}$ [44], $Pb^{2+}$, 1-naphthol and 1-naphthylamine [45], etc. In addition, we found at lower pH (pH ~ 2), the alginate/alginic component was heavily inhibited of, but GO became highly contributed when comparing their performances of the beads with/without GO. This could attribute to that GO covered multiple interactions with AO, including that of oxygen functional groups having pH dependent dissociation, though [43] .

## 4. Conclusions

GO-encapsulated alginate and porous alginic beads were prepared at the viewpoint of environmental consciousness. With characterization, the results showed GO was well incorporated and also the beads were promoted more porous than those in the absence of GO. Kinetic studies demonstrated that the addition of GO shortened the adsorption time, raised the initial adsorption rate and uplifted the adsorption capacity. Isotherm studies indicated the adsorptions followed Langmuir-type, and the GO-containing beads had comparably high capacities. In addition, this study provided a low-cost route for GO/alginate-based application in water decontamination.

**References**

[1] Zhao G, Li J, Ren X, Chen C, Wang X. Few-layered graphene oxide nanosheets as superior sorbents for heavy metal ion pollution management. Environ Sci Technol 2011;45(24):10454-62.

[2] Zhao GX, Ren XM, Gao X, Tan XL, Li JX, Chen CL, et al. Removal of Pb(II) ions from aqueous solutions on few-layered graphene oxide nanosheets. Dalton Trans 2011;40(41):10945-52.

[3] Gao Y, Li Y, Zhang L, Huang H, Hu J, Shah SM, et al. Adsorption and removal of tetracycline antibiotics from aqueous solution by graphene oxide. J Colloid Interface Sci 2012;368(1):540-6.

[4] Liu S, Tian J, Wang L, Luo Y, Sun X. One-pot synthesis of CuO nanoflower-decorated reduced graphene oxide and its application to photocatalytic degradation of dyes. Catal Sci Technol 2012;2(2):339.

[5] Fan L, Luo C, Sun M, Li X, Lu F, Qiu H. Preparation of novel magnetic chitosan/graphene oxide composite as effective adsorbents toward methylene blue. Bioresour Technol 2012;114:703-6.

[6] Sun HM, Cao LY, Lu LH. Magnetite/reduced graphene oxide nanocomposites: One step solvothermal synthesis and use as a novel platform for removal of dye pollutants. Nano Res 2011;4(6):550-62.

[7] Wang K, Ruan J, Song H, Zhang JL, Wo Y, Guo SW, et al. Biocompatibility of graphene Oxide. Nanoscale Res Lett 2011;6(1):8.

[8] Schinwald A, Murphy FA, Jones A, MacNee W, Donaldson K. Graphene-based nanoplatelets: a new risk to the respiratory system as a consequence of their unusual aerodynamic properties. ACS Nano 2011;6(1):736-46.

[9] Papageorgiou SK, Katsaros FK, Kouvelos EP, Nolan JW, Le Deit H, Kanellopoulos NK. Heavy metal sorption by calcium alginate beads from Laminaria digitata. J Hazard Mater 2006;137(3):1765-72.

[10] Park HG, Chae MY. Novel type of alginate gel-based adsorbents for heavy metal removal. J Chem Technol Biotechnol 2004;79(10):1080-3.

[11] Tam NFY, Wong YS, Simpson CG. Repeated removal of copper by alginate beads and the enhancement by microalgae. Biotechnol Tech 1998;12(3):187-90.

[12] Sartori C, Finch DS, Ralph B, Gilding K. Determination of the cation content of alginate thin films by FTi.r. spectroscopy. Polymer 1997;38(1):43-51.

[13] Kuhn SP, Pfister RM. Adsorption of mixed metals and cadmium by calcium-alginate immobilized Zoogloea ramigera. Appl Microbiol Biotechnol 1989;31(5):613-8.

[14] Ai L, Zhang C, Chen Z. Removal of methylene blue from aqueous solution by a solvothermal-synthesized graphene/magnetite composite. J Hazard Mater 2011;192(3):1515-24.

[15] Rocher V, Siaugue JM, Cabuil V, Bee A. Removal of organic dyes by magnetic alginate beads. Water Res 2008;42(4-5):1290-8.

[16] Lin YB, Fugetsu B, Terui N, Tanaka S. Removal of organic compounds by alginate gel beads with entrapped activated carbon. J Hazard Mater 2005;120(1-3):237-41.

[17] Fugetsu B, Satoh S, Shiba T, Mizutani T, Lin YB, Terui N, et al. Caged multiwalled carbon nanotubes as the adsorbents for affinity-based elimination of ionic dyes. Environ Sci Technol 2004;38(24):6890-6.

[18] Sone H, Fugetsu B, Tanaka S. Selective elimination of lead(II) ions by alginate/polyurethane composite foams. J Hazard Mater 2009;162(1):423-9.

[19] He Y, Zhang N, Gong Q, Qiu H, Wang W, Liu Y, et al. Alginate/graphene oxide fibers with enhanced mechanical strength prepared by wet spinning. Carbohydr Polym 2012;88(3):1100-8.

[20] Sun L, Yu H, Fugetsu B. Graphene oxide adsorption enhanced by in situ reduction with sodium hydrosulfite to remove acridine orange from aqueous solution. J Hazard Mater 2012;203-204(0):101-10.

[21] Hummers WS, Offeman RE. Preparation of graphitic oxide. J Am Chem Soc 1958;80(6):1339.

[22] Fugetsu B, Satoh S, Iles A, Tanaka K, Nishi N, Watari F. Encapsulation of multi-walled carbon nanotubes (MWCNTs) in $Ba^{2+}$-alginate to form coated micro-beads and their application to the pre-concentration/elimination of dibenzo-p-dioxin, dibenzofuran, and biphenyl from contaminated water. Analyst 2004;129(7):565-6.

[23] Qiu H, Lv L, Pan B, Zhang Q, Zhang W, Zhang Q. Critical review in adsorption kinetic models. J Zhejiang Univ Sci A 2009;10(5):716-24.

[24] Ho YS, McKay G. Pseudo-second order model for sorption processes. Process Biochem 1999;34(5):451-65.

[25] Hameed BH, El-Khaiary MI. Malachite green adsorption by rattan sawdust: isotherm, kinetic and mechanism modeling. J Hazard Mater 2008;159(2-3):574-9.

[26] Zheng H, Liu DH, Zheng Y, Liang SP, Liu Z. Sorption isotherm and kinetic modeling of aniline on Cr-bentonite. J Hazard Mater 2009;167(1-3):141-7.

[27] Li LB, Fang YP, Vreeker R, Appelqvist I. Reexamining the egg-box model in calcium-alginate gels with X-ray diffraction. Biomacromolecules 2007;8(2):464-8.

[28] Yang G, Zhang L, Peng T, Zhong W. Effects of $Ca^{2+}$ bridge cross-linking on structure and pervaporation of cellulose/alginate blend membranes. J Membr Sci 2000;175(1):53-60.

[29] Hu B, Fugetsu B, Yu H, Abe Y. Prussian blue caged in spongiform adsorbents using diatomite and carbon nanotubes for elimination of cesium. J Hazard Mater 2012;217-218:85-91.

[30] Fugetsu B, Sano E, Sunada M, Sambongi Y, Shibuya T, Wang XS, et al. Electrical conductivity and electromagnetic interference shielding efficiency of carbon nanotube/cellulose composite paper. Carbon 2008;46(9):1256-8.

[31] Kim J, Cote LJ, Kim F, Yuan W, Shull KR, Huang JX. Graphene oxide sheets at interfaces. J Am Chem Soc 2010;132(23):8180-6.

[32] Zhang NN, Qiu HX, Si YM, Wang W, Gao JP. Fabrication of highly porous biodegradable monoliths strengthened by graphene oxide and their adsorption of metal ions. Carbon 2011;49(3):827-37.

[33] Poncelet D, Lencki R, Beaulieu C, Halle JP, Neufeld RJ, Fournier A. Production of alginate beads by emulsification/internal gelation. I. Methodology. Appl Microbiol Biotechnol 1992;38(1):39-45.

[34] Poncelet D, Poncelet De Smet B, Beaulieu C, Huguet ML, Fournier A, Neufeld RJ. Production of alginate beads by emulsification/internal gelation. II. Physicochemistry. Appl Microbiol Biotechnol 1995;43(4):644-50.

[35] Papageorgiou SK, Kouvelos EP, Favvas EP, Sapalidis AA, Romanos GE, Katsaros FK. Metal-carboxylate interactions in metal-alginate complexes studied with FTIR spectroscopy. Carbohydr Res 2010;345(4):469-73.

[36] Valentin R, Horga R, Bonelli B, Garrone E, Di Renzo F, Quignard F. Acidity of alginate aerogels studied by FTIR spectroscopy of probe molecules. Macromol Symp 2005;230(1):71-7.

[37] Qin YM, Hu HQ, Luo AX. The conversion of calcium alginate fibers into alginic acid fibers and sodium alginate fibers. J Appl Polym Sci 2006;101(6):4216-21.

[38] Konishi Y, Asai S, Shimaoka J, Miyata M, Kawamura T. Recovery of neodymium and ytterbium by biopolymer gel particles of alginic acid. Ind Eng Chem Res 1992;31(10):2303-11.

[39] Dogan H. Preparation and characterization of calcium alginate-based composite adsorbents for the removal of Cd, Hg, and Pb ions from aqueous solution. Toxicol Environ Chem 2012;94(3):482-99.

[40] Lv G, Li Z, Jiang W-T, Chang P-H, Jean J-S, Lin K-H. Mechanism of acridine orange removal from water by low-charge swelling clays. Chem Eng J 2011;174(2-3):603-11.

[41] Chattopadhyay S, Traina SJ. Spectroscopic study of sorption of nitrogen heterocyclic compounds on phyllosilicates. Langmuir 1999;15(5):1634-9.

[42] Rocher V, Bee A, Siaugue JM, Cabuil V. Dye removal from aqueous solution by magnetic alginate beads crosslinked with epichlorohydrin. J Hazard Mater 2010;178(1-3):434-9.

[43] Zhou CJ, Zhang WJ, Zhou WC, Lei AH, Zhang QL, Wan Q, et al. Fast and considerable adsorption of methylene blue dye onto graphene oxide. Bull Environ Contam Toxicol 2011;87(1):86-90.

[44] Mi X, Huang G, Xie W, Wang W, Liu Y, Gao J. Preparation of graphene oxide aerogel and its adsorption for $Cu^{2+}$ ions. Carbon 2012;50(13):4856-64.

[45] Yang X, Chen C, Li J, Zhao G, Ren X, Wang X. Graphene oxide-iron oxide and reduced graphene oxide-iron oxide hybrid materials for the removal of organic and inorganic pollutants. Rsc Adv 2012;2(23):8821.